\RequirePackage{fix-cm}
\documentclass[onecolumn,epjc3]{svjour3}  
\smartqed  
\RequirePackage{graphicx}
\usepackage{multicol}
\usepackage{multirow}
\usepackage{tikz,xcolor,hyperref}
\RequirePackage{float}
\RequirePackage{hyperref}
\RequirePackage{amsmath}
\RequirePackage{amssymb}
\usepackage{widetext}
\RequirePackage{mathtools}  
\usepackage{caption}
\definecolor{lime}{HTML}{A6CE39}
\DeclareRobustCommand{\orcidicon}{%
	\begin{tikzpicture}
	\draw[lime, fill=lime] (0,0) 
	circle [radius=0.20] 
	node[white] {{\fontfamily{qag}\selectfont \tiny ID}};
	\draw[white, fill=white] (-0.0625,0.095) 
	circle [radius=0.012];
	\end{tikzpicture}
	\hspace{-2mm}
}

\foreach \x in {A, ..., Z}{%
	\expandafter\xdef\csname orcid\x\endcsname{\noexpand\href{https://orcid.org/\csname orcidauthor\x\endcsname}{\noexpand\orcidicon}}
}

\journalname{Eur. Phys. J. C}
\begin{document}

\title{Growth Rate and Configurational Entropy in Tsallis Holographic Dark Energy}


\author{Snehasish Bhattacharjee \orcidA{} \thanksref{e1,addr1}                
      }

\thankstext{e1}{e-mail: snehasish.bhattacharjee.666@gmail.com}


\institute{Department of Physics, Indian Institute of Technology, Hyderabad 502285, India  \label{addr1}           
     }

\date{\today}

\maketitle

\begin{abstract}
In this work, we analyzed the effect of different prescriptions of the IR cutoffs, namely the Hubble horizon cutoff, particle horizon cutoff, Granda and Oliveros horizon cut off, and the Ricci horizon cutoff on the growth rate of clustering for the Tsallis holographic dark energy (THDE) model in an FRW universe devoid of any interactions between the dark Universe. Furthermore, we used the concept of configurational entropy to derive constraints (qualitatively) on the model parameters for the THDE model in each IR cutoff prescription from the fact that the rate of change of configurational entropy hits a minimum at a particular scale factor $a_{DE}$ which indicate precisely the epoch of dark energy domination predicted by the relevant cosmological model as a function of the model parameter(s). By using the current observational constraints on the redshift of transition from a decelerated to an accelerated Universe, we derived constraints on the model parameters appearing in each IR cutoff definition and on the non-additivity parameter $\delta$ characterizing the THDE model and report the existence of simple linear dependency between $\delta$ and $a_{DE}$ in each IR cutoff setup.     

\end{abstract}

\section{Introduction}

The dynamics of the Universe is well explained by the $\Lambda$CDM cosmological model, albeit with the presumption that the Universe contains dark matter and dark energy in overwhelming quantities \cite{1to4} where the former dictates how the objects in the Universe cluster and the latter on how should the Universe expand. Much of modern cosmology has therefore been focused on pinpointing the properties of these quantities to understand the dynamical evolution of the Universe in the past and the future. In this spirit, many dedicated experiments have been proposed or are in operation but have so far not yielded any conclusive evidence for the existence of the dark Universe and therefore critically affect the viability of the standard cosmological model. Additionally, the existence of the so-called Hubble tension \cite{5to8} makes things worse and hints at the possible existence of unknown physics and on the possibility of the Universe being governed by an entirely different cosmological model. It, therefore, compels us to explore other alternatives to explain these profound cosmological enigmas. \\
Holographic dark energy (HDE) models are proving to be viable alternatives to the static cosmological constant in addressing some of the above problems. Constructed upon the holographic principle \cite{hooft,susskind,cohen}, HDE models are also in harmony with the latest observational results \cite{zhang,zhang2,huong,enqvist,shen}. In HDE, the horizon entropy plays the most crucial role which if changed, changes the dynamics of the model significantly \cite{notes}. \\
Exercising the modified entropy-area relation reported in \cite{tsallis} and in conjunction with the holographic principle, the authors in \cite{tavayef} proposed the Tsallis HDE which could explain the observed accelerated expansion of the universe and has been widely studied in the literature. For example, in \cite{notes}, the authors studied the dynamics of THDE for various IR cutoff scales and also for non-interacting and interacting scenarios. In \cite{bhatta}, the dynamics of THDE have been studied for the hybrid expansion law. In \cite{bdt}, THDE model was used to investigate late-time cosmology in Brans-Dicke gravity. In \cite{branet}, the cosmological features of THDE were studied in Cyclic, DGP, and RS II braneworlds. Stability analysis concerning the THDE model with an interaction between dark matter and dark energy was studied in \cite{st}. In \cite{obt}, cosmological perturbations in the linear regime on subhorizon scales for the THDE model assuming a future event horizon as the IR cutoff was investigated and constraints on model parameters were imposed. Interested readers can refer to \cite{wang} for a recent review on various HDE models. \\
As first proposed in \cite{19}, the cosmic evolution from a near-perfect gaussian matter distribution to the highly non-linear state owing to the clustering of dark matter haloes could be due to the dissipation of configurational entropy. Additionally, the depletion of configurational entropy could also be responsible for the currently observed cosmic expansion \cite{20}. The depletion of configurational entropy is inevitable for a static Universe, since it is unstable due to the presence of large-scale structures. If we assume the Universe as a whole to comprise of a thermodynamic system, it, therefore, must obey the second law of thermodynamics which warrants cosmic expansion to suppress the growth of structure formation due to the absence of entropy generation processes to check the dissipation of configurational entropy \cite{20}. It must also be noted that the configurational entropy dissipates for a dust Universe and only damps out in an accelerating Universe and that the first derivative of configurational entropy attains a minimum at a scale factor corresponding to the epoch of the dark energy domination for that model \cite{20}.\\
The concept of configurational entropy has been used to study cosmology in power-law $f(T)$ gravity \cite{ft} and in Chaplygin gas models \cite{cg}. In this work, we shall investigate the growth rate of clustering in the THDE model for various IR cutoffs to understand how does the different IR cutoffs influence the growth rate concerning the standard $\Lambda$CDM model and also to put qualitative constraints on the model parameters from the first derivative of the configurational entropy. \\
The manuscript is organized as follows: In Section \ref{sec2}, we briefly summarize the concept of configurational entropy and growth rate. In Section \ref{sec4}, we discuss the Tsallis holographic dark energy model in different IR cutoffs and discuss the results and in Section \ref{sec6} we conclude. Throughout the work we use $\Omega_{m0}=0.315$, $\Omega_{\Lambda}=1-\Omega_{m0}$ and $h=0.674$ \cite{planck} and work with natural units.

\section{Configurational entropy and growth rate }\label{sec2} 
Let us consider a large comoving volume $V$ in the cosmos where the assumptions of isotropy and homogeneity hold true. Now, $V$ can be divided into a number of smaller volume elements $dV$ with energy density $\rho (\overrightarrow{x},t)$ where $\overrightarrow{x}$ and $t$ represent respectively the comoving coordinates and time.\\
Now, from the definition of configurational entropy in \cite{19} which was further motivated from the definition of information entropy in \cite{shannon}, we can write
\begin{equation}
\Phi (t) = - \int \rho \hspace{0.05cm}\text{log} \hspace{0.05cm} \rho d V .
\end{equation}
Next, the equation of continuity for an expanding Universe can be expressed as  
\begin{equation}\label{eq1}
 3 \frac{\dot{a}}{a} \rho +\frac{\partial \rho}{\partial t}+ \frac{1}{a} \bigtriangledown . (\rho \overrightarrow{\nu})  = 0,
\end{equation}
where $a$ and $\overrightarrow{\nu}$ represent respectively the scale factor and the peculiar velocity of cosmic fluid in $dV$. \\
Next, multiplying Eq. \ref{eq1} with $(1+ \text{log} \rho)$, and then integrating over volume $V$, we arrive at the following differential equation \cite{19} 
\begin{equation}\label{1}
\frac{d \Phi(a)}{d a} \dot{a}+ 3 \frac{\dot{a}}{a} \Phi (a) - \chi (a)  = 0,
\end{equation}
where $\chi (a) = \frac{1}{a} \int \rho (\overrightarrow{x},a) \bigtriangledown . \overrightarrow{\nu} d V$. \\
Now, the divergence of peculiar velocity $\nu$ can be written as \cite{20} 
\begin{equation}\label{2}
\bigtriangledown.\nu(\overrightarrow{x}) = -a \dot{a} \frac{d D (a)}{d a} \delta (\overrightarrow{x}),
\end{equation}
where $D(a)$ is the growing mode of fluctuations and $\delta (\overrightarrow{x})$ represents the density contrast at a given position $\overrightarrow{x}$. Upon substituting Eq. \ref{2} in Eq. \ref{1}, yields after some calculation \cite{20}
\begin{equation}\label{3}
\frac{d \Phi (a)}{d a} +  \frac{3}{a}( \Phi (a)-\Psi) + \overline{\rho} f    \frac{ D^{2} (a)}{ a} \int\delta^{2}(\overrightarrow{x})dV  = 0,
\end{equation}
where $\Psi = \int \rho (\overrightarrow{x},a) dV  $ defines the total mass enclosed within the volume $V$ and $\overline{\rho}$ represents the critical density of the cosmos and 
\begin{equation}\label{growth}
f = \frac{d \text{ln}D}{d \text{ln}a}=  \Omega_{m}(a)^{\gamma}
\end{equation} 
represents the dimensionless growth rate with $\Omega_{m}(a)$ being the matter density parameter and $\gamma$ being the growth index. The growth index in Einstein's gravity is roughly equal to 6/11 \cite{stein} while it is slightly different for modified gravity models and also for dynamical dark energy models. Ref \cite{frt}  derived an analytical expression for $\gamma$ in $f(R,T)$ theory of gravity (where $R$ denote the Ricci scalar and $T$ denote the trace of the Energy-Momentum tensor) which could be used to constrain the model parameters for that particular class of modified gravity models. Different parametrizations for $\gamma$ exist in the literature which are both constant and redshift dependent. In this work, let us work with \cite{linder/index}
\begin{equation}\label{index}
\gamma = 0.55 + 0.05 (1+\omega (a=0.5)).
\end{equation}  \\
Solution of Eq. \ref{3} provides the evolution of configurational entropy in the cosmological model under consideration. In order to numerically solve Eq. \ref{3}, we assume the time independent quantities in Eq. \ref{3} to be equal to 1 and set the initial condition $\Phi(a_{i}) = \Psi$.\\
The interesting feature of Eq. \ref{3} is that the first derivative of configurational entropy (i.e, $\frac{d\Phi(a)}{da}$) attains a minima at a particular scale factor $a_{DE}$ after which the dark energy domination takes place \cite{20}. The location of $a_{DE}$ for a particular cosmological model is solely dependent on the relative supremacy of the dark energy and its influence on the growth history of cosmic structures \cite{20}. It is evident from the third term of Eq. \ref{3} that the rate of change of configurational entropy depends on the precise conjunction of the scale factor, growing mode, and its time derivative and it is this distinct combination of these quantities which is accountable for the well-defined minimum observed in the rate of change of configurational entropy, and denote the epoch of an accelerating Universe dominated by dark energy, and is a novel technique to constrain the model parameters of any cosmological model. So, by comparing the minima obtained from the $\Lambda$CDM model with that obtained from other dark energy models, we can put qualitative constraints on the model parameters of different dark energy models and this is precisely the motivation for this work.

\section{Tsallis holographic dark energy}\label{sec4}
\subsection{Expression of energy density}
The modified black hole's horizon entropy suggested in \cite{tsallis} takes the form $S_{\delta}=\gamma D^{\delta}$, where $\delta$ represent the Tsallis or the non-additivity parameter, $\gamma$ a constant and $D$ the surface area of the Black hole's event horizon and for $\delta=1$, the usual Bekenstein entropy can be recovered. \\
The relationship between the entropy $S$, IR and UV cutoffs denoted respectively by $\Pi$ and $\Theta$ is given by \cite{hooft,susskind,cohen}
\begin{equation}
S^{3/4}\geq (\Pi\Theta)^{3}.
\end{equation} 
Now, since the energy density for the HDE models scales as $\rho_{DE}\propto \Theta^{4}$, the general expression for the energy density of the THDE can be written as \cite{tavayef}
\begin{equation}\label{main}
\rho_{DE}=A \Pi^{2\delta-4},
\end{equation}
where $A$ is an unknown parameter. Considering a flat FRW background with $'-','+','+','+'$ metric signature, the first Friedmann equation assumes the form
\begin{equation}\label{eq4}
H^{2}=\frac{1}{3}\left( \rho_{DE} + \rho_{m}\right) 
\end{equation}
where $\rho_{m}$ and $\rho_{DE}$ are the energy density of the dark matter and the dark energy respectively and therefore the corresponding dimensionless density parameters are defined as
\begin{equation}
\Omega_{DE}=\frac{\rho_{DE}}{\rho_{c}}= \frac{A}{3}H^{2-2\delta}, \hspace*{0.25in} \Omega_{m}=\frac{\rho_{m}}{\rho_{c}},
\end{equation}
with $\rho_{c}=3H^{2}$ being the critical density of the Universe. \\
It can be clearly seen from Eq. \ref{main} that different choices for the IR cutoffs will generate different expressions for the THDE energy density with significantly different dark energy EoS parameter $\omega_{DE}$. There are currently four different prescriptions for the IR cutoffs, namely the Hubble horizon cutoff, particle horizon cutoff \cite{par}, Granda and Oliveros horizon cutoff \cite{go} and the Ricci horizon cutoff \cite{ricci}. In the subsequent subsections, we shall numerically compute the growth rate and the rate of change of configurational entropy for the THDE model with the aforementioned IR cutoffs. 

\subsection{Hubble horizon cutoff} 
For the standard HDE model, the Hubble horizon does not give rise to an accelerated expansion \cite{par}. However, as shown in \cite{tavayef}, for the THDE model, an accelerated expansion is possible for such a cutoff even in the absence of interacting scenarios.\\
Therefore, for the first case, we shall assume the IR cutoff to be the Hubble horizon represented by 
\begin{equation}\label{eq.hub}
\Pi = \frac{1}{H}.
\end{equation}
Substituting Eq. \ref{eq.hub} in Eq. \ref{main}, the expression of $\rho_{DE}$ takes the form
\begin{equation}\label{rho.hub}
\rho_{DE}=A H^{4-2\delta}.
\end{equation}
The evolution of the dark energy density parameter ($\Omega_{DE}$) can be obtained by numerically solving the following differential equation \cite{tavayef}
\begin{equation}\label{den.hub}
\Omega^{'}_{DE}=3\Omega_{DE}(\delta-1)\left[\frac{1-\Omega_{DE}}{1-\Omega_{DE}(2-\delta)} \right],
\end{equation} 
where $\Omega^{'}_{DE} = d \Omega^{'}_{DE} / d \text{ln}a$. Finally, the EoS parameter ($\omega_{DE}$) reads \cite{tavayef}
\begin{equation}\label{omega.hub}
\omega_{DE}=\frac{\delta-1}{\Omega_{DE}(2-\delta)-1}.
\end{equation} 
Now, in order to compute the growth rate ($f(a)$), we substitute Eq. \ref{omega.hub} into Eq. \ref{index} and Eq. \ref{rho.hub} into Eq. \ref{growth} since $\Omega_{m}(a)=1-\Omega_{DE}(a)$ and then substitute the result in Eq. \ref{3} to compute the rate of change of configurational entropy.   
\begin{figure}[H]
\centering
  \includegraphics[width=7.5 cm]{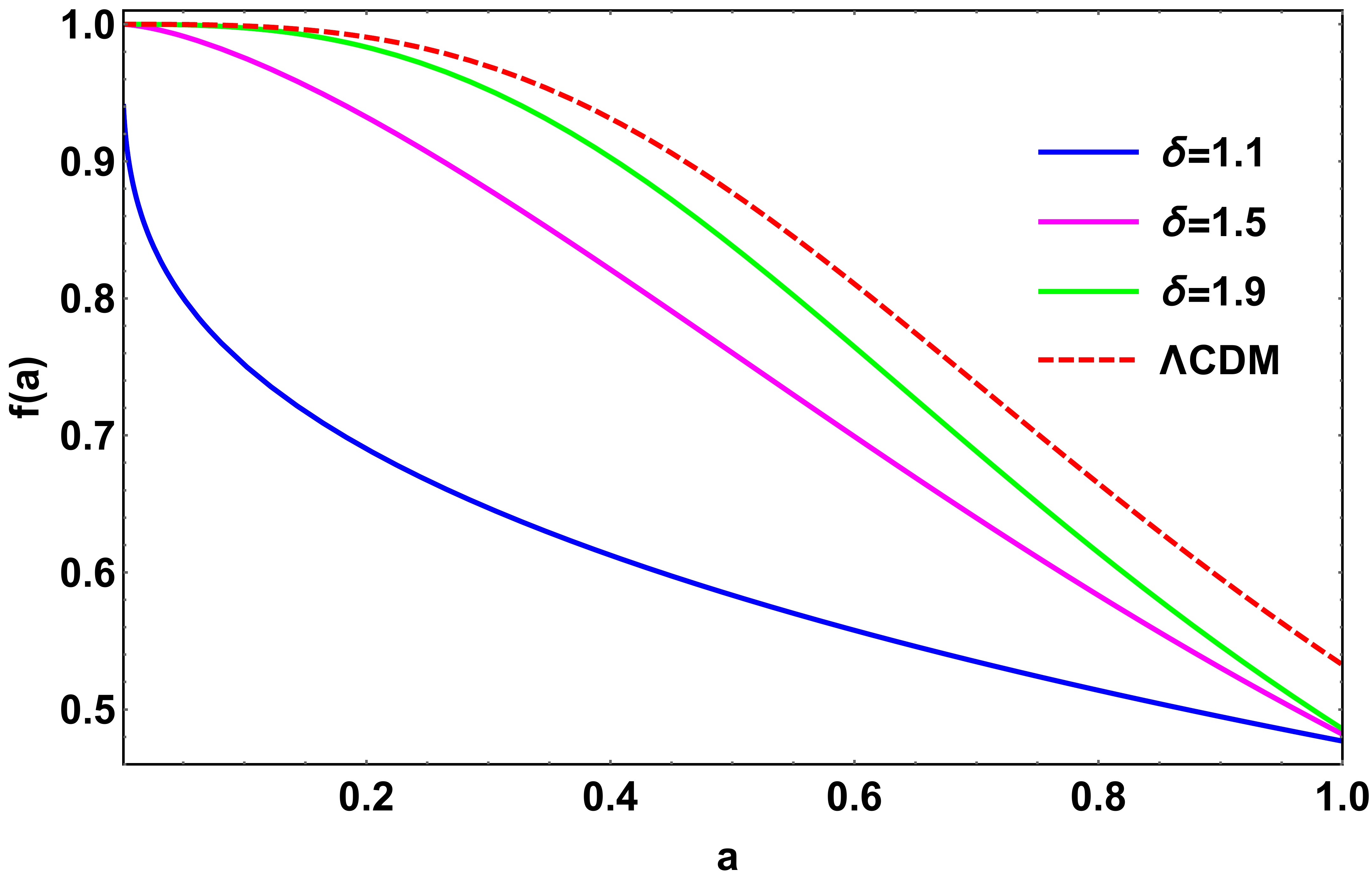}  
  \includegraphics[width=8.4 cm]{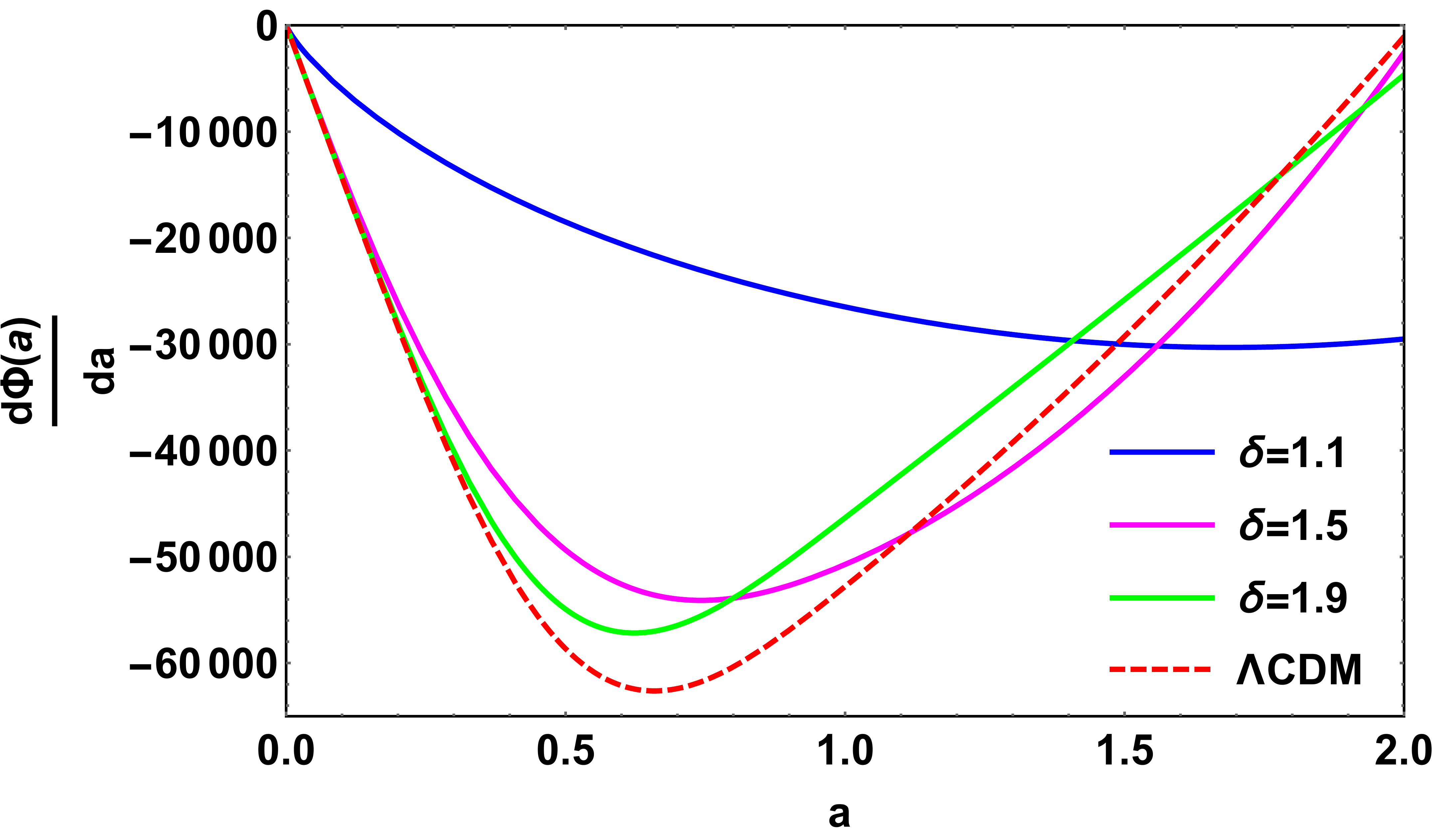}
  \includegraphics[width=8.4 cm]{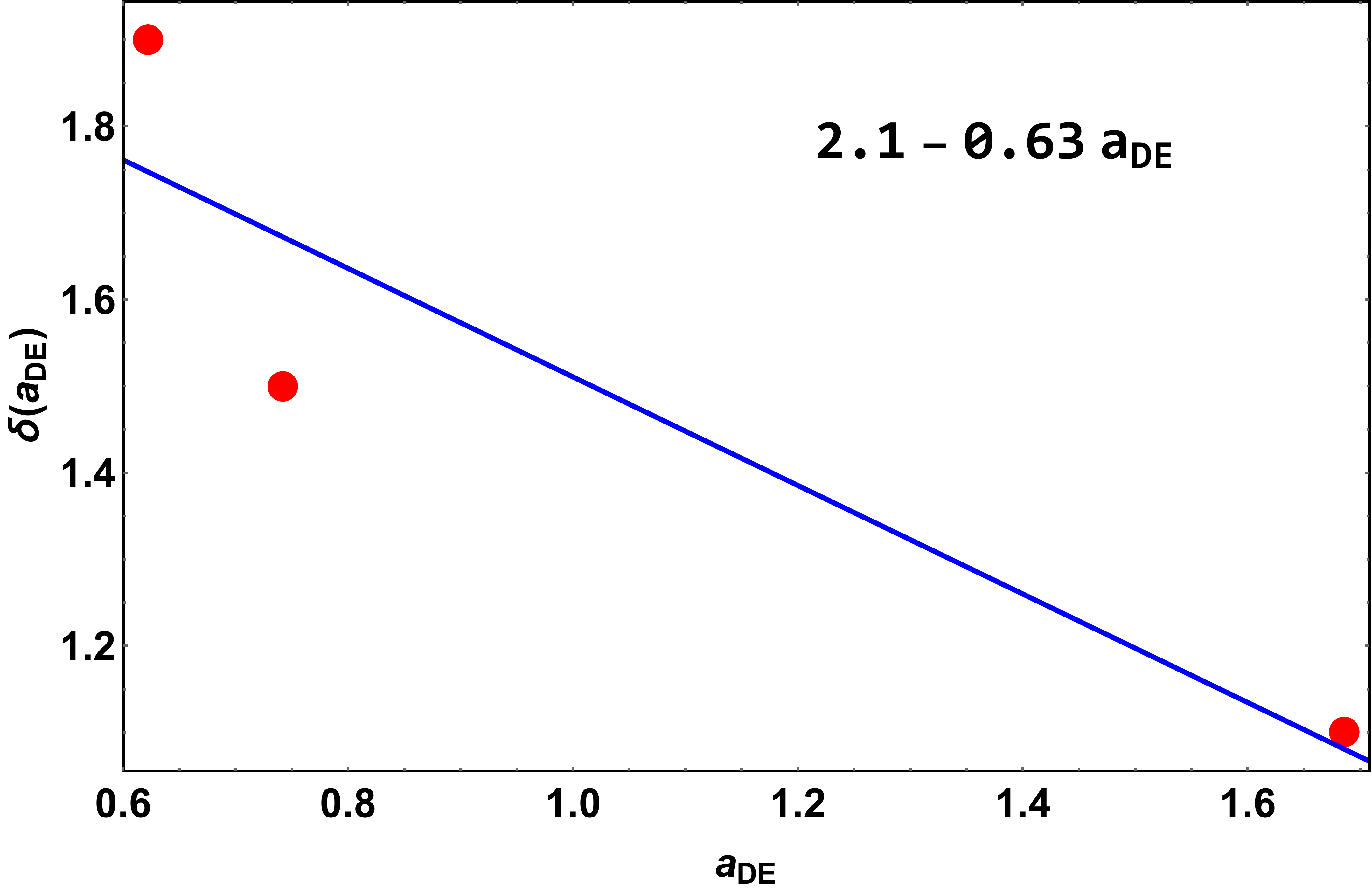}
\caption{Top left panel shows the evolution of growth rate $f(a)$, top right panel shows the rate of change of configurational entropy ($\frac{d\Phi(a)}{da}$), and the lower panel shows the best fit relation between the non-additivity parameter $\delta$ and the scale factor $a_{DE}$ at which $\frac{d\Phi(a)}{da}$ attains a minima.}
\label{FIG1}
\end{figure}
From the plot of $f(a)$, it is clear that the suppression of growth rate is minimum for the $\Lambda$CDM model, and the steepness increases as the non-additivity parameter $\delta$ decreases. However, the profiles start to coincide at low redshifts (high scale factor). Additionally, as the parameter $\delta$ approaches two, the profiles become indistinguishable from the $\Lambda$CDM model at all times. \\
From the $\frac{d\Phi(a)}{da}$ plot, a clear dissipation of configurational entropy is seen in all cases with the $\Lambda$CDM model showing the maximum dissipation. This can be explained by the fact that configurational entropy is a derived quantity and largely depends on the magnitude of the growth rate. Therefore, since the $\Lambda$CDM model exhibits the maximum growth rate of clustering, it also tends to dissipate the highest amount of configurational entropy and the scale factor of minima occurs at $a_{DE}\simeq0.65$ which corresponds to a redshift of $z_{DE}\simeq0.53$. For the THDE model, this dissipation, as usual, depends on the non-additivity parameter $\delta$, with reduced dissipation rates as $\delta$ is lowered. For $\delta=1.9$, the scale factor of minima occurs at $a_{DE}=0.62$ which corresponds to a redshift of $z_{DE}=0.612$,  while for the $\delta=1.5$ and $1.1$ cases, $a_{DE}\simeq1.69$  $(z_{DE}=-0.41)$ and 0.74 $(z_{DE}=0.35)$ respectively. The $\delta=1.1$ case is unphysical and therefore cannot be considered a viable option. For $\delta=1.9$ and transition occur at a redshift which is very close to that obtained in the $\Lambda$CDM scenario and is also consistent with current observational bounds \cite{bound}. The $\delta=1.5$ case predicts an accelerating Universe at a much higher redshift which is again observationally unfavorable.\\
In the bottom panel, the plot between $\delta$ and $a_{DE}$ is shown where a moderate linear dependency is observed with the best fit relation of $\delta=-0.63 a_{DE} + 2.1$. 
\subsection{Particle horizon cutoff}
In the second case, we set the IR cutoff to be the particle horizon defined as \cite{par} 
\begin{equation}\label{eq.par}
R_{p} = a(t)\int_{0}^{t} \frac{dt}{a(t)},
\end{equation}
which satisfies the equation 
\begin{equation}
\dot{R}_{p}=1+ H R_{p}.
\end{equation}
The expression of the energy density ($\rho_{DE}$) in this setup reads \cite{notes}
\begin{equation}\label{rho.par}
\rho_{DE}=AR_{p}^{-4+2\delta},
\end{equation}
The evolution of $\Omega_{DE}$ in this setup can be obtained by solving the following differential equation \cite{notes}
\begin{equation}
\Omega^{'}_{DE} = \Omega_{DE}(\Omega_{DE}-1)\left( 2 \Sigma (\delta-2) + 1 - 2\delta\right), 
\end{equation}
where 
\begin{equation}
\Sigma = \left[\frac{3 \Omega_{DE} H^{-2+2\delta}}{A} \right] ^{\frac{1}{-2\delta+4}},
\end{equation}
and the expression for the EoS parameter ($\omega_{DE}$) reads \cite{notes}
\begin{equation}
\omega_{DE}= -\left[1 + \left(\frac{2\delta-4}{3} \right)\left(1+\Sigma \right)  \right]. 
\end{equation}

\begin{figure}[H]
\centering
  \includegraphics[width=7.5 cm]{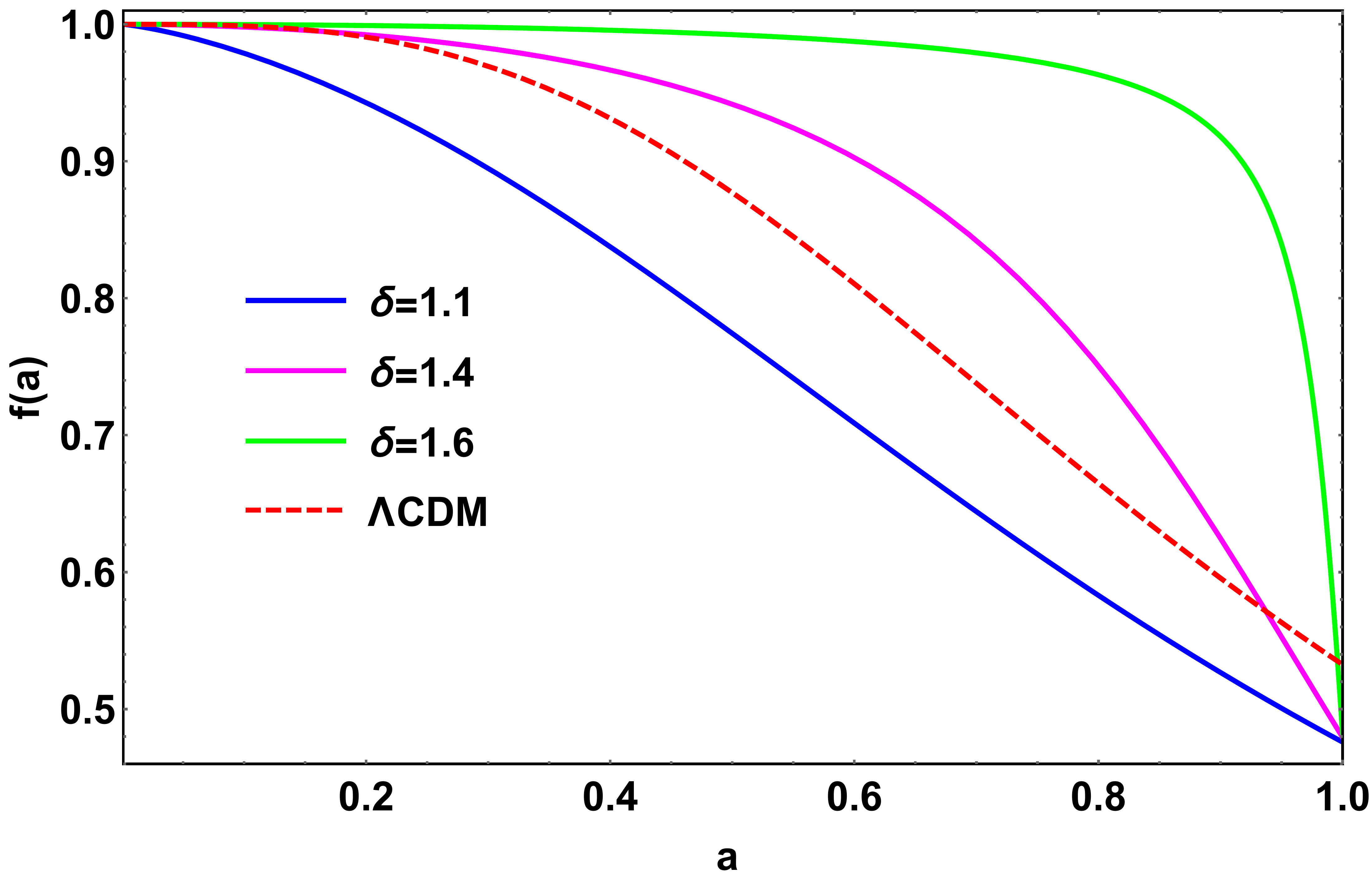}  
  \includegraphics[width=8.4 cm]{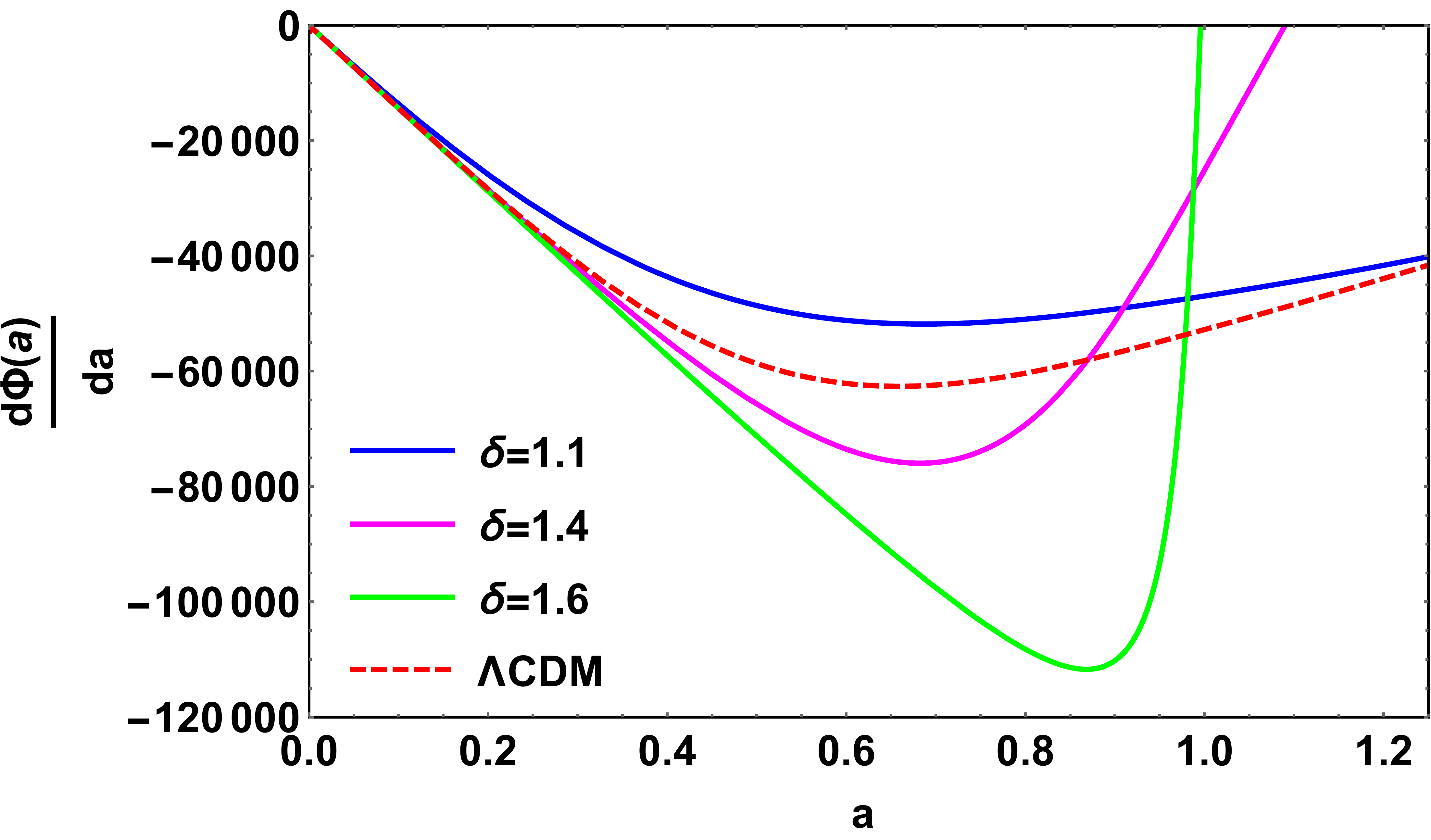}
  \includegraphics[width=8.4 cm]{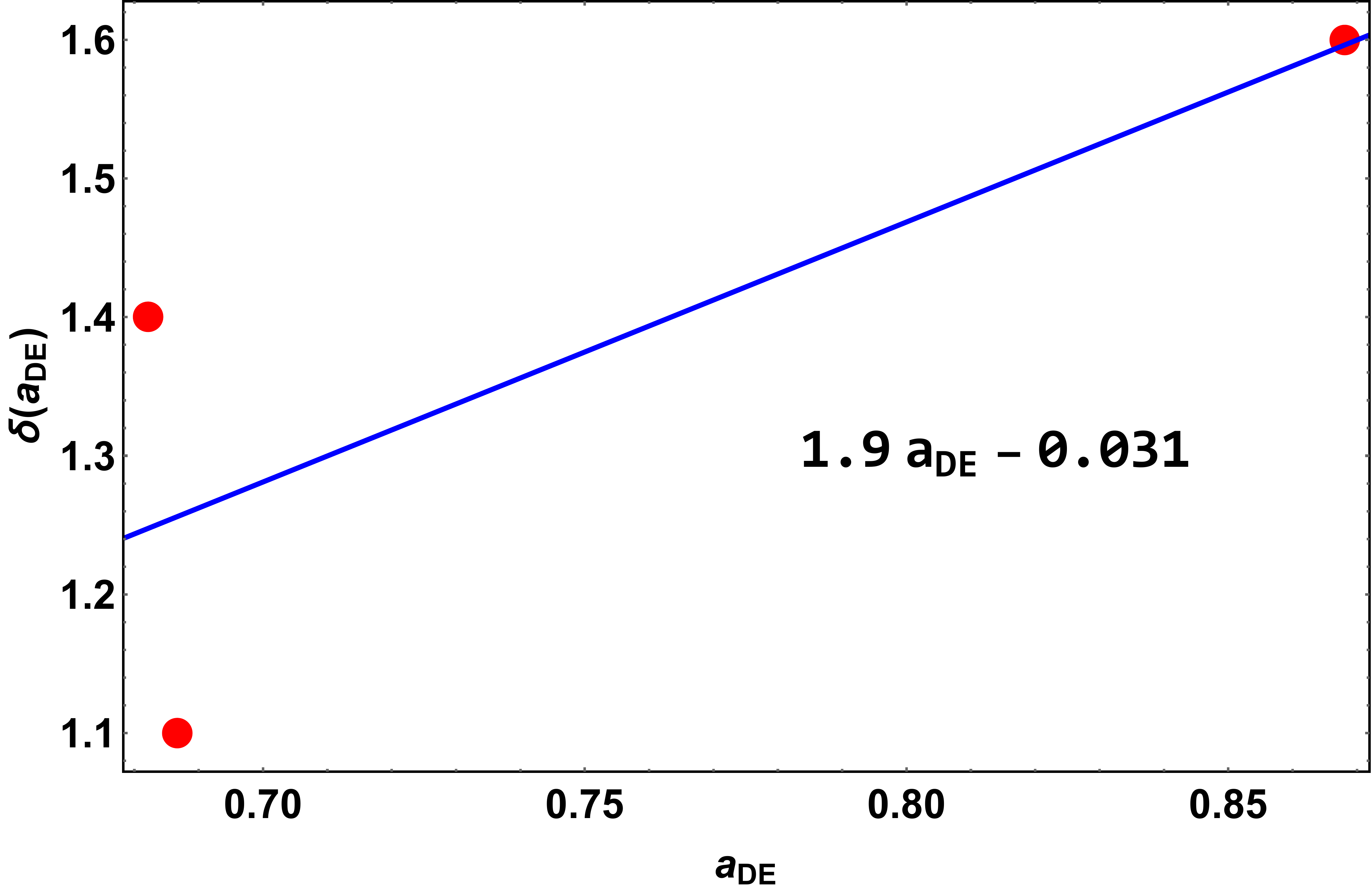}
\caption{Top left panel shows the evolution of growth rate $f(a)$, top right panel shows the rate of change of configurational entropy ($\frac{d\Phi(a)}{da}$), and the lower panel shows the best fit relation between the non-additivity parameter $\delta$ and the scale factor $a_{DE}$ at which $\frac{d\Phi(a)}{da}$ attains a minima. The plots are drawn for $A=10$.}
\label{FIG2}
\end{figure}
The plot of growth rate $f(a)$ for the particle horizon cutoff depicts the $\Lambda$CDM model to have a significantly lower growth rate than the $\delta=1.6$ and 1.4 cases, while the $\delta=1.1$ case has the least $f(a)$ and has the steepest slope. The $\delta=1.6$ case deviate the most from the $\Lambda$CDM model.\\
From the $\frac{d\Phi(a)}{da}$ plot, the $\delta=1.6$ case as expected exhibit the maximum dissipation of configurational entropy with the minima occurring at $a_{DE}=0.87$ $ (z_{DE}=0.15)$ which is observationally not consistent \cite{bound}. However, for the $\delta=1.4$ and $\delta=1.1$ cases, the minima occur respectively at   $a_{DE}=0.68$ $(z_{DE}=0.45)$ and $a_{DE}=0.68$ $(z_{DE}=0.466)$ which is very close to the $\Lambda$CDM model with the best fit relation between $\delta$ and $a_{DE}$ reading $\delta = 1.9 a_{DE} -0.031$.
\subsection{Granda and Oliveros (GO) horizon cutoff}
 The GO cutoff was proposed by Granda and Oliveros \cite{go} to resolve the causality and coincidence problems in cosmology and is defined as $\Pi = (m H^{2} + n \dot{H})^{-1/2}$ for which the expression of the energy density takes the form \cite{notes}
 \begin{equation}
 \rho_{DE} = (m H^{2} + n \dot{H})^{2-\delta}.
 \end{equation}
The differential equation required to solve for the evolution of $\Omega_{DE}$ and the expression of the EoS parameter ($\omega_{DE}$) reads respectively as \cite{notes}
\begin{equation}
\Omega^{'}_{DE}=(1-\Omega_{DE})\left[ \frac{2}{n}\left(-m+\frac{(3\Omega_{DE})^{\frac{1}{2-\delta}}}{H^{\frac{2-2\delta}{2-\delta}}} \right)+3 \right], 
\end{equation} 
and
\begin{equation}
\omega_{DE}=-1-\frac{1}{3\omega_{DE}}\left[3(1-\Omega_{DE})+\frac{2}{n}\left(-m+\frac{(3\Omega_{DE})^{\frac{1}{2-\delta}}}{H^{\frac{2-2\delta}{2-\delta}}} \right) \right].
\end{equation}
\begin{figure}[H]
\centering
  \includegraphics[width=7.5 cm]{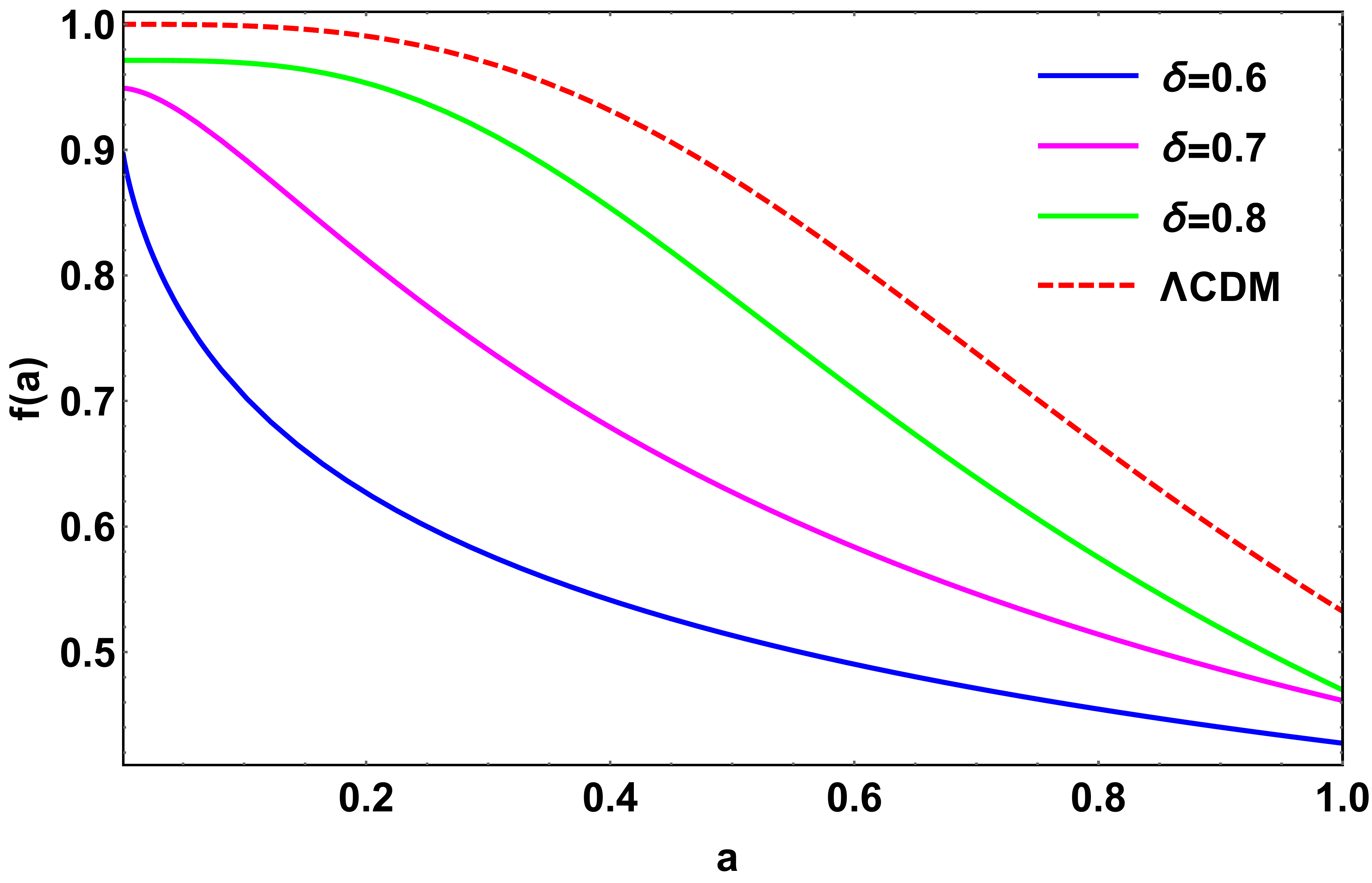}  
  \includegraphics[width=8.4 cm]{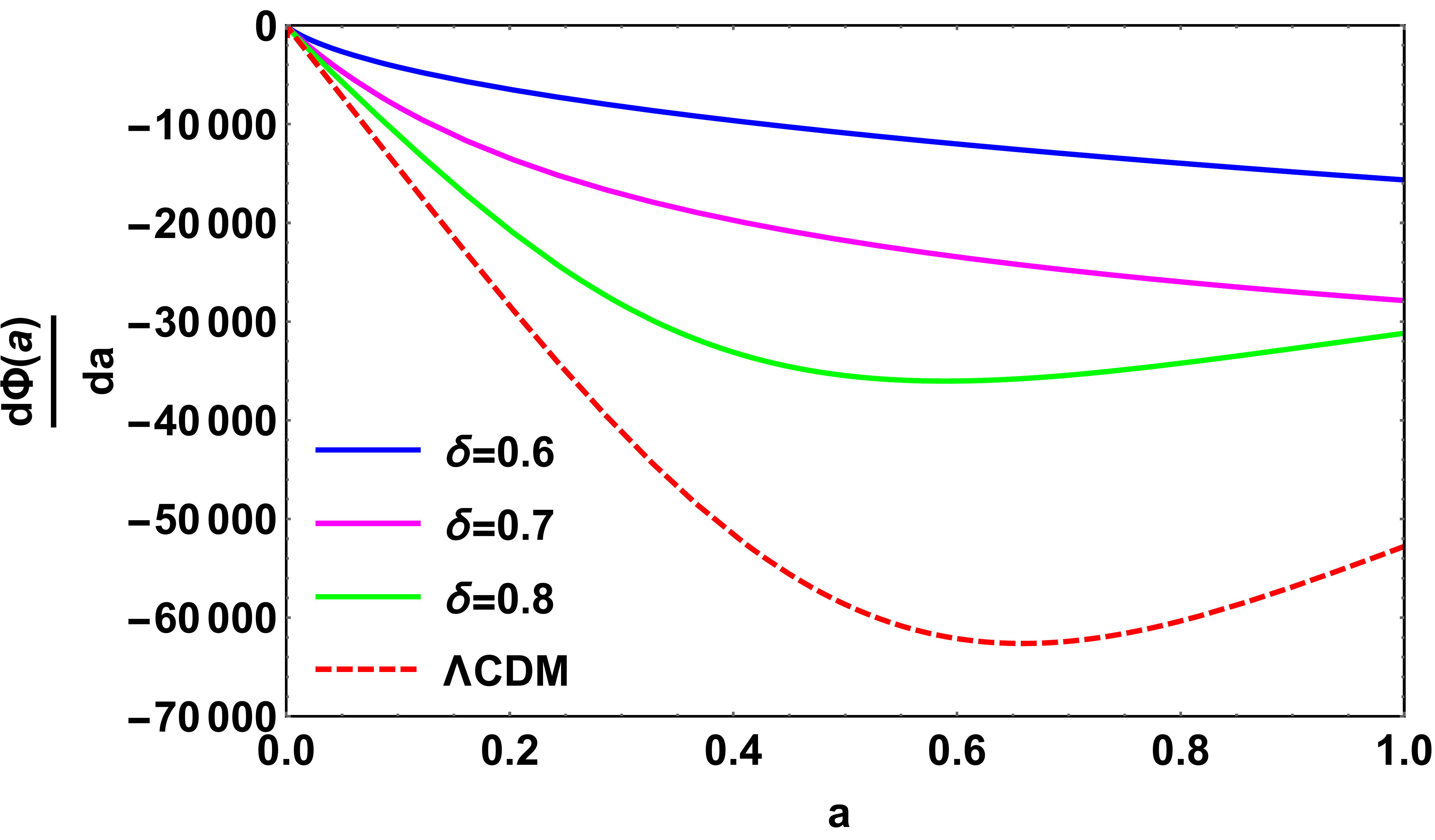}
  \includegraphics[width=8.4 cm]{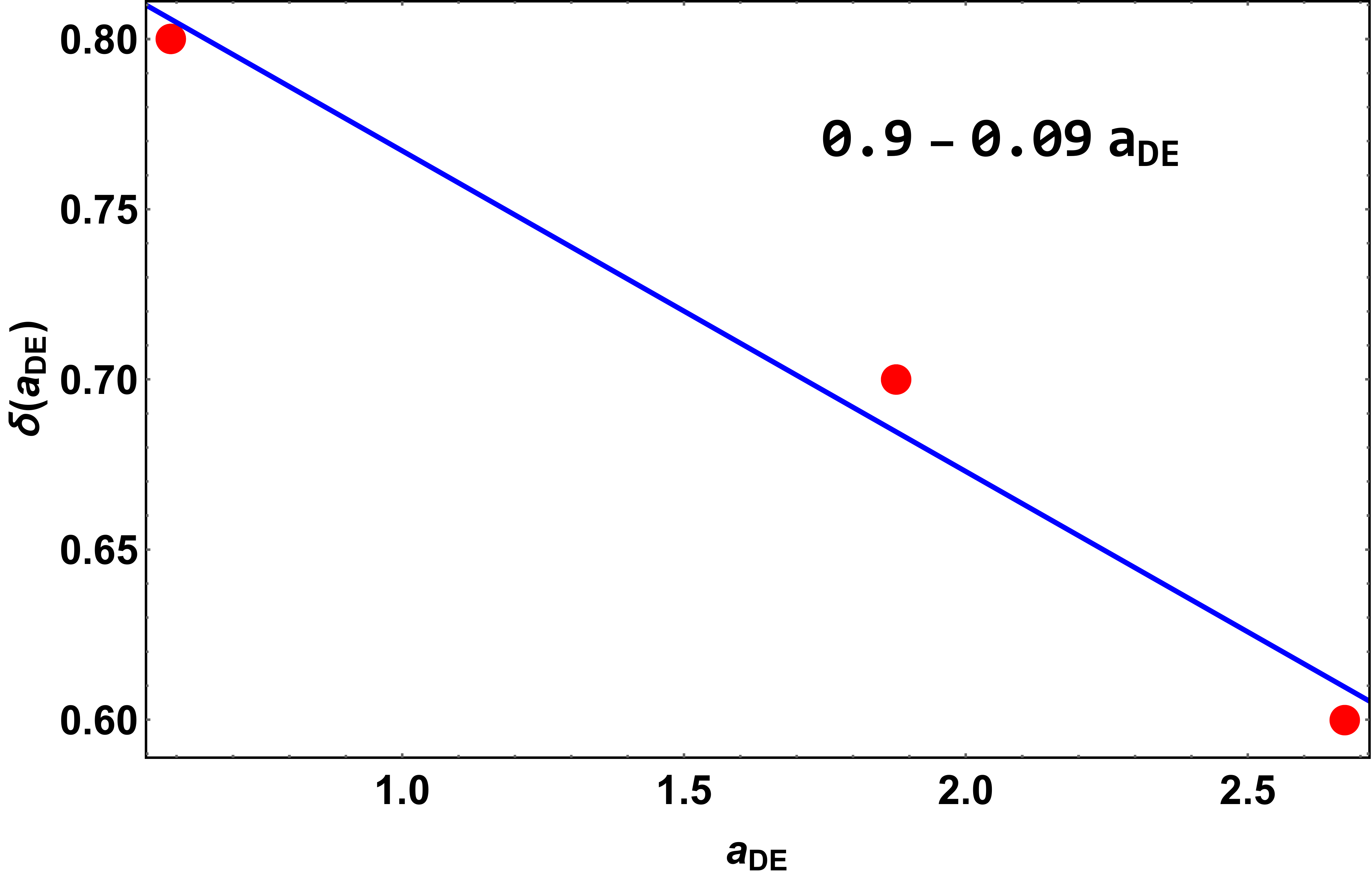}
\caption{Top left panel shows the evolution of growth rate $f(a)$, top right panel shows the rate of change of configurational entropy ($\frac{d\Phi(a)}{da}$), and the lower panel shows the best fit relation between the non-additivity parameter $\delta$ and the scale factor $a_{DE}$ at which $\frac{d\Phi(a)}{da}$ attains a minima. The plots are drawn for $m=0.8$ and $n=0.5$.}
\label{FIG3}
\end{figure}
The growth rate $f(a)$ for the THDE model with the GO cutoff differs substantially from the $\Lambda$CDM model. The growth rate diminishes rapidly as $\delta$ is lowered. It is also evident that in all cases $f(a)$ is lower than that of the $\Lambda$CDM model.\\
The $\frac{d\Phi(a)}{da}$ plot shows the $\Lambda$CDM model to dispel maximum configurational entropy similar to the case when the IR cutoff is modeled with the Hubble horizon. For $\delta=0.6$ and $0.7$, the minima occurs at $a_{DE}=2.67$ $(z_{DE}=-0.62)$ and $1.88  (z_{DE}=-0.46)$ respectively, both of which predict a decelerating Universe and therefore unphysical. Nonetheless, for $\delta=0.8$, the minima occurs at $a_{DE}=0.59$ $(z_{DE}=0.68)$ which falls well within the current observational constraints \cite{bound} and also very close to the same obtained from the $\Lambda$CDM model. A strict linear dependency between $\delta$ and $a_{DE}$ is observed with the best fit relation $\delta = -0.09 a_{DE} + 0.9$. \\
In \cite{notes} the authors used the same combination of the free parameters (i.e, $m$, $n$ and $\delta$) and reported viable estimates of kinematical parameters in all cases. However, from this work, it is clear that only for $\delta=0.8$, an accelerating universe is possible. Therefore, it is apparent that the rate of change of configurational entropy provides an alternative method to constrain the free parameters and also furnish a robust consistency check to the constraints obtained from other methods and statistics.  
\subsection{Ricci horizon cutoff}

For the fourth and final case, we shall set the Ricci horizon as the IR cutoff for which the expression of the THDE model reads \cite{ricci,notes}
\begin{equation}
\rho_{DE}=\lambda\left(\dot{H} + 2 H^{2} \right) ^{2-\delta}
\end{equation}
where $\lambda$ is an unknown parameter as usual \cite{par}. Similar to the previous case, the first order differential equation which generate the evolution of $\Omega_{DE}$ and the expression of the EoS parameter ($\omega_{DE}$) reads respectively as \cite{notes}
\begin{equation}
\Omega^{'}_{DE}=(1-\Omega_{DE})\left[2\left(-2+\frac{(3\Omega_{DE} \lambda^{-1})\frac{1}{2-\delta}}{H^{\frac{2-2\delta}{2-\delta}}} \right)  \right], 
\end{equation}
and 
\begin{equation}
\omega_{DE}=-1-\left( \frac{1-\Omega_{DE}}{\Omega_{DE}}\right) \left[1+\frac{2}{3(1-\Omega_{DE})} \left(-2+\frac{(3\Omega_{DE} \lambda^{-1})\frac{1}{2-\delta}}{H^{\frac{2-2\delta}{2-\delta}}} \right)\right]. 
\end{equation}
\begin{figure}[H]
\centering
  \includegraphics[width=7.5 cm]{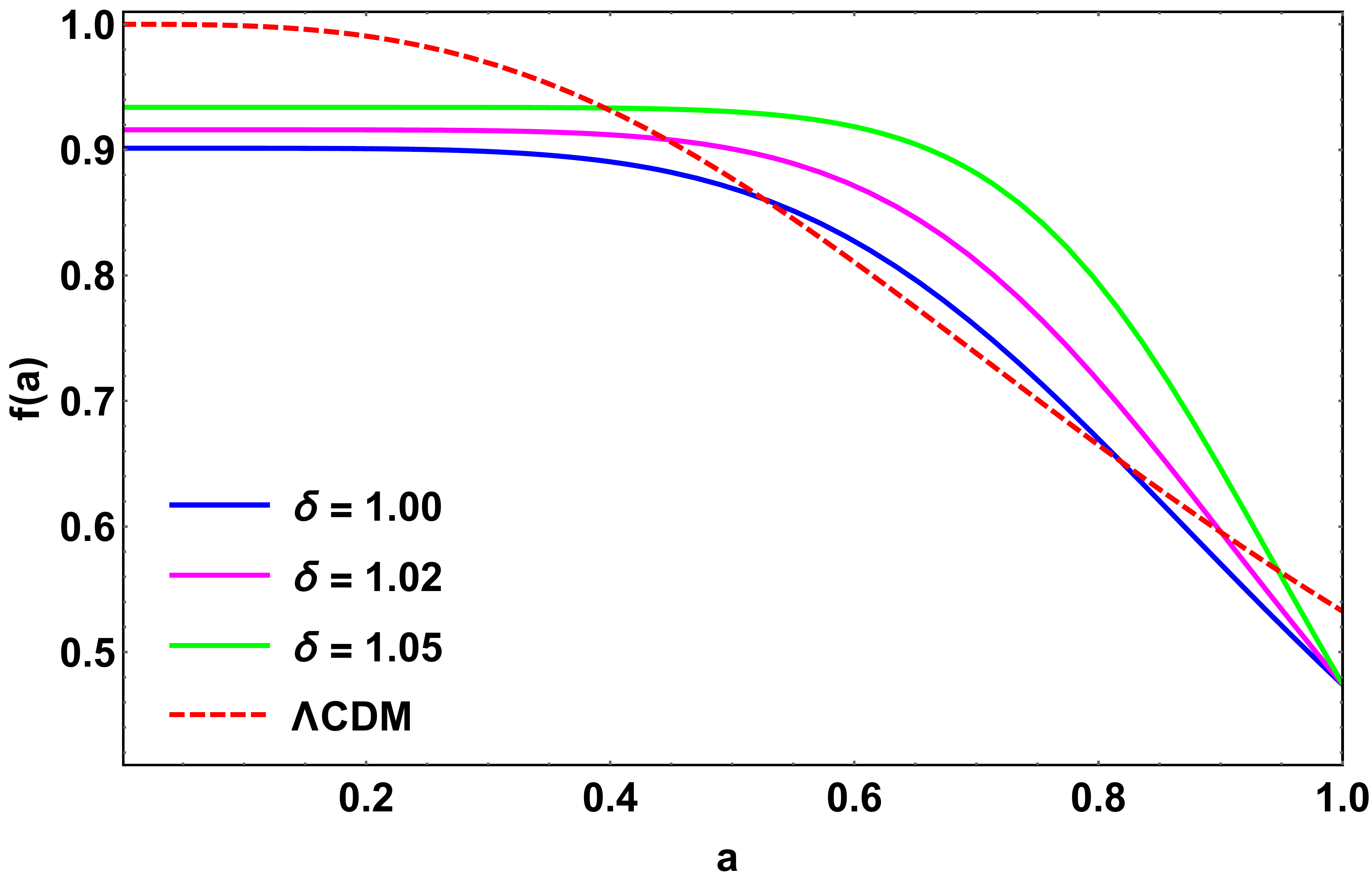}  
  \includegraphics[width=8.4 cm]{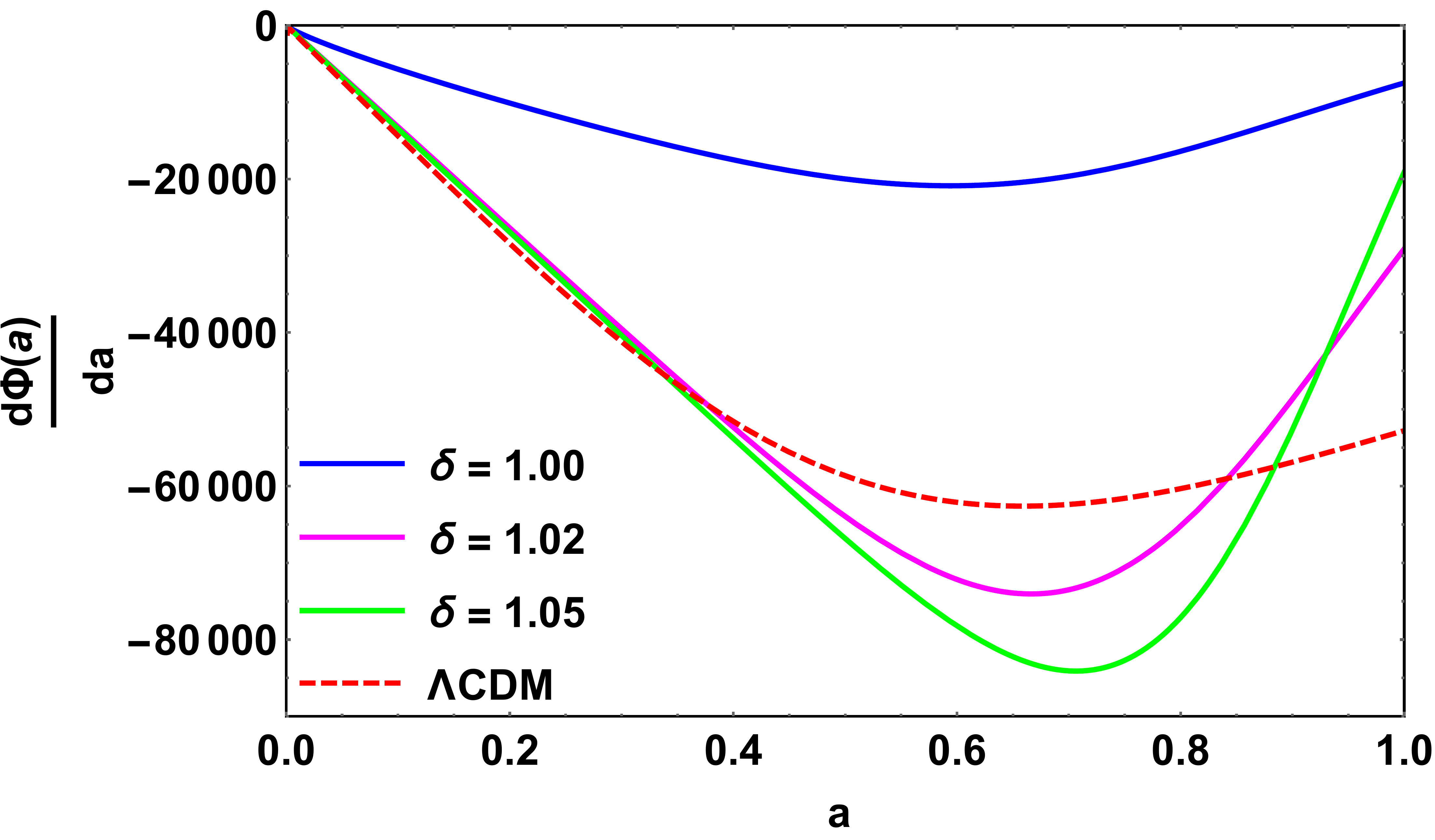}
  \includegraphics[width=8.4 cm]{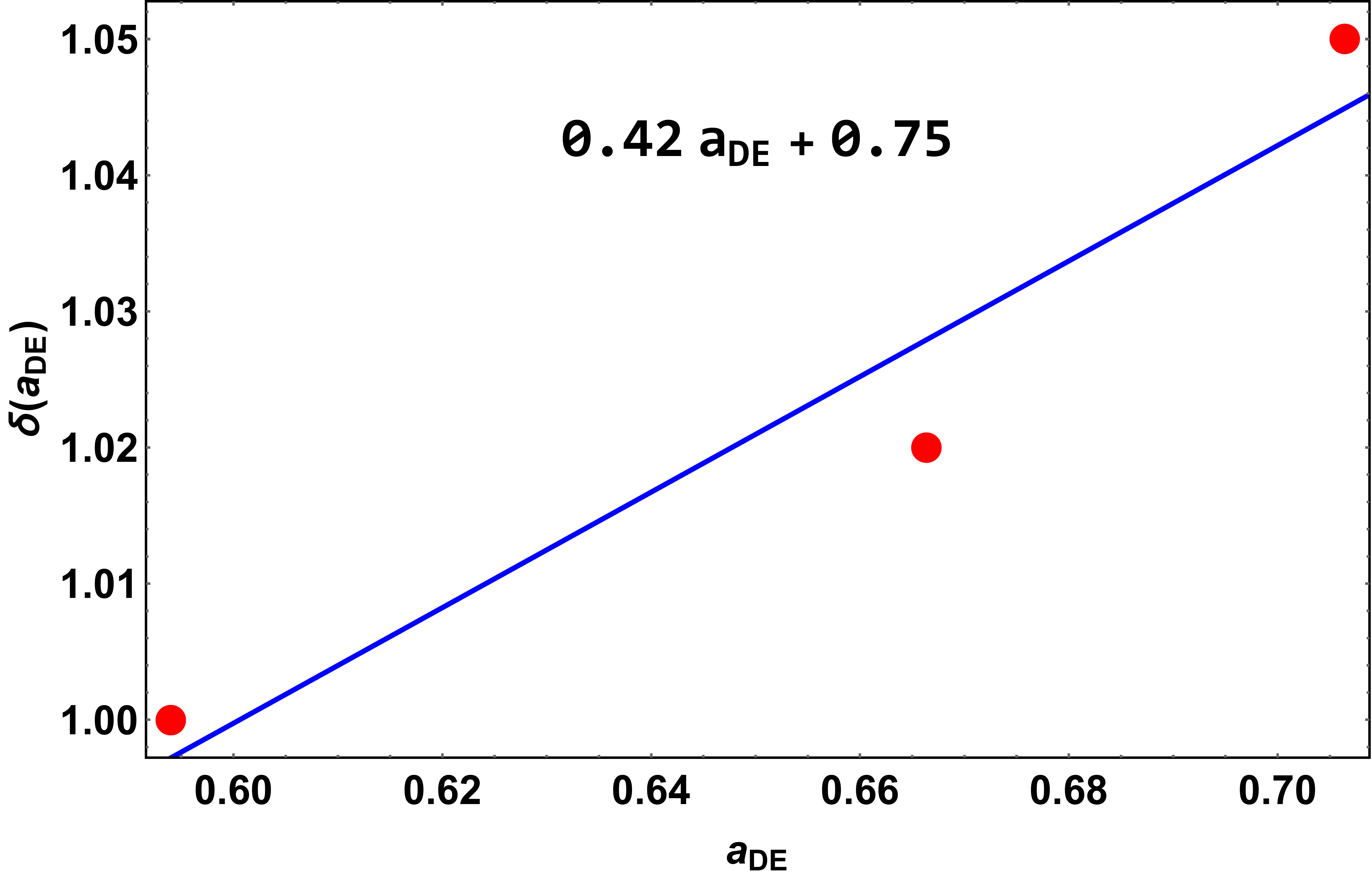}
\caption{Top left panel shows the evolution of growth rate $f(a)$, top right panel shows the rate of change of configurational entropy ($\frac{d\Phi(a)}{da}$), and the lower panel shows the best fit relation between the non-additivity parameter $\delta$ and the scale factor $a_{DE}$ at which $\frac{d\Phi(a)}{da}$ attains a minima. The plots are drawn keeping $\lambda$ unity.}
\label{FIG4}
\end{figure}
The $f(a)$ plot in this case shows the $\Lambda$CDM profile to predict significantly higher growth rate than the THDE model prior a specific scale factor dependent solely on the non-additivity parameter $\delta$ following which the growth rate in the THDE model exceeds the one for the $\Lambda$CDM. For $\delta=1$, the profile intersects the $\Lambda$CDM at $a\simeq0.52$ and at $a\simeq0.81$, for $\delta=1.02$, the crossing occurs at $a\simeq0.44$ and $a\simeq0.9$ and for $\delta=1.05$, the same transpires at $a\simeq0.39$ and $a\simeq0.94$. It is also noticeable that profiles tend to be steeper with decreasing $\delta$. \\
From the $\frac{d\Phi(a)}{da}$ plot, the minima for the $\delta=1.0, 1.02$ and $1.05$ cases transpire at $a_{DE}\simeq0.59, 0.66,$ and $0.71$ which in terms of redshift reads $z_{DE}=0.69, 0.51,$ and $0.42$ respectively. The $\delta=1.02$ case predict the redshift of transition in harmony with observations while for the $\delta=1$ and  $1.05$ cases, the theoretical prediction of $z_{DE}$ is unsupported with respect to the current observations constraints \cite{bound}. A clear linear relation between $\delta$ and $a_{DE}$ materializes with the best fit relation $\lambda=0.42 a_{DE} + 0.75$.
 
\section{Conclusions}\label{sec6}
Holographic dark energy (HDE) models are a promising alternative to the static cosmological constant in explaining the late-time dynamics of the Universe and have already alleviated some of the major hitches plaguing the $\Lambda$CDM cosmological model \cite{zhang,zhang2,huong,enqvist,shen}. These models are built upon the holographic principle \cite{hooft,susskind,cohen} where the horizon entropy plays the most crucial role with different IR cutoffs predicting completely different dynamics for the HDE model under consideration \cite{notes}.  \\
Ref \cite{20} reported that the first derivative of configurational entropy attains a minimum at a particular scale factor $a_{DE}$ after which the dark energy domination takes place.  The location of $a_{DE}$ for a particular cosmological model is solely dependent on the relative supremacy of the dark energy and its influence on the growth history of cosmic structures \cite{20}. It is evident from the third term of Eq. \ref{3} that the rate of change of configurational entropy depends on the exact conjunction of the scale factor, growing mode, and its time derivative and it is this distinct combination of these quantities which is accountable for the well-defined minimum observed in the rate of change of configurational entropy, and denote the epoch of an accelerating Universe dominated by dark energy, and is a novel technique to constrain the model parameters of any cosmological model. \\
Thus, upon employing the idea, we tried to constrain the non-additivity parameter $\delta$ for the THDE model and other free parameters appearing in each IR cutoff case by juxtaposing the theoretical estimate of $a_{DE}$ in each case with that reported by current observations.\\ 
We find that there exists a suitable parameter range between which the THDE model predicts an accelerating universe in each IR cutoff recipe at a suitable redshift consistent with observations and report the existence of simple linear dependencies between the non-additivity parameter $\delta$ and $a_{DE}$ in each IR cutoff prescription.\\
It may be noted that in most of the previous studies related to the THDE model, the model parameters were constrained by obtaining observationally consistent values for some of the cosmological parameters such as the deceleration parameter, EoS parameter, and Hubble parameter, while in this paper, we have imposed constraints on the model parameters from the fact that the rate of change of configurational entropy must exhibit a minimum at a certain observationally favored scale factor range which denotes the epoch of an accelerating Universe, and therefore present an alternative method to constrain the model parameters for the THDE model and also furnish a robust consistency check to the constraints obtained from other methods and statistics.\\
The motivation to carry out the analysis with multiple IR cutoffs was simply to check whether all of these IR cutoff prescriptions lead to the rate of change of configurational entropy attaining the minimum at the suitable scale factor range consistent with observations, and understand how does the growth rate of clustering differs in each IR cutoff recipe when compared with the standard $\Lambda$CDM model, and finally to check whether any dependency is observed between the scale factor of minimum ($a_{DE}$) and the non-additivity parameter ($\delta$). Now, since there exist corners in parameter spaces for all the IR cutoff prescriptions, we do not report here the existence of a more realistic IR cutoff. Thorough investigations are required in this regard to compare the usability, and understand which of these IR cutoffs best describes the dynamics of the Universe.\\
As a final note we add that although the THDE model is turning out to be promising, more work is needed to understand its viability in cosmology. We plan to carry out an analysis to constrain the non-additivity parameter $\delta$ from the constraints coming from Big-Bang nucleosynthesis and check whether the constraints obtained could explain the accelerated expansion. Such a study could turn out to be an acid test for the THDE model.

\newpage
\begingroup
\setlength{\tabcolsep}{10pt} 
\renewcommand{\arraystretch}{2} 
\begin{tabular}{ | p{3cm}||p{1cm}|p{1cm}| p{1cm}||p{4cm}|  }

 \hline
 IR cutoffs &$\delta$ & $a_{DE}$
 &$z_{DE}$ & Best fit equation\\
 \hline
\multirow{3}{*} { Hubble horizon}   &   1.1 & 0.74 &  0.35 & \\ \cline{2-4}   &   1.5 & 1.69 &  -0.41 & $\delta=-0.63a_{DE}+2.1$\\ \cline{2-4}   &   1.9 & 0.62 &  0.612 &\\
 \hline
  \hline
 \multirow{3}{*} { Particle horizon}   &   1.1 & 0.68 &  0.466 & \\ \cline{2-4}   &   1.4 & 0.68 &  0.45 & $\delta=1.9a_{DE}-0.031$\\ \cline{2-4}   &   1.6 & 0.87 &  0.15 &\\
 \hline
  \hline
 \multirow{3}{*} { GO horizon}   &   0.6 & 2.67 &  -0.62 & \\ \cline{2-4}   &   0.7 & 1.88 &  -0.46 & $\delta=-0.09a_{DE}+0.9$\\ \cline{2-4}   &   0.8 & 0.59 &  0.68 &\\
 \hline
  \hline
\multirow{3}{*} { Ricci horizon}   &   1.00 & 0.59 &  0.69 & \\ \cline{2-4}   &   1.02 & 0.66 &  0.51 & $\delta=0.42a_{DE}+0.75$\\ \cline{2-4}   &   1.05 & 0.71 &  0.42 &\\
 \hline
 \hline 
\end{tabular}
\captionof{table}{The table shows the scale factor for minimum $a_{DE}$, redshift of transition $z_{DE}$ for different choices of the non-additivity parameter $\delta$ and the best fit equation for $\delta$ as a function of $a_{DE}$ for the THDE model with different IR cutoff prescriptions.
}\label{table2}
\endgroup

\section*{Acknowledgments}
I thank an anonymous reviewer for constructive criticisms and useful suggestions and Shantanu Desai for reading the manuscript and for the fruitful discussions. I also thank DST, New-Delhi, Government of India for the provisional INSPIRE fellowship selection [Code: DST/INSPIRE/03/2019/003141].

\end{document}